\documentclass[pra,twocolumn,showpacs,superscriptaddress,floatfix]{revtex4}
\usepackage[dvipdfmx,dvipdfmx]{graphicx,color}
\usepackage{subfigure}
\graphicspath{{./Figure/}}
\usepackage{epsfig}
\usepackage{float}

\usepackage{amsmath}	
\newcommand{\abs}[1]{\left| #1 \right|}
\newcommand{\ket}[1]{\left | #1 \right \rangle}
\def\k(#1){|#1\rangle}
\newcommand{\bra}[1]{\left \langle #1 \right |}
\newcommand{\proj}[1]{\ket{#1} \bra{#1}}

\newcommand{\tr}{{\rm \, Tr }\, }

\newcommand{\beq}{\begin{equation}}
\newcommand{\eeq}{\end{equation}}
\newcommand{\beqa}{\begin{eqnarray}}
\newcommand{\eeqa}{\end{eqnarray}}
\newcommand{\beqan}{\begin{eqnarray*}}
\newcommand{\eeqan}{\end{eqnarray*}}

\newcommand{\affA}{%
\affiliation{
 National Institute of Information and Communications Technology,
 4-2-1 Nukui-kita, Koganei, Tokyo 184-8795, Japan}
     }
\newcommand{\affB}{%
\affiliation{
 Sophia University,
 7-1 Kioicho, Chiyoda-ku, Tokyo 102-8554, Japan}
}

\begin{document}

\title{\bf Optical phase estimation via coherent state 
and displaced-photon counting
}


\author{Shuro Izumi }
\affA \affB
\author{Masahiro Takeoka}%
\affA
\author{Kentaro Wakui}%
\affA
\author{Mikio Fujiwara}%
\affA
\author{Kazuhiro Ema}%
\affB
\author{Masahide Sasaki}%
\affA
%


\begin{abstract}


We consider the phase sensing via weak optical coherent state at quantum limit precision. 
A new detection scheme for the phase estimation is proposed which is inspired by the suboptimal quantum measurement in coherent optical communication. 
We theoretically analyze a performance of our detection scheme, which we call the displaced-photon counting, for phase sensing in terms of the Fisher information and show that the displaced-photon counting outperforms the static homodyne and heterodyne detections in wide range of the target phase. 
The proof-of-principle experiment is performed with linear optics and a superconducting nanowire single photon detector. 
The result shows that our scheme overcomes the limit of the ideal homodyne measurement even under practical imperfections.

\end{abstract}

\maketitle

\section{Introduction}\label{Sect:1}
Quantum optical sensing has attracted both fundamental and practical 
interests (see for example \cite{DS15} and references therein). 
In particular, sensing of unknown phase $\phi$ of the optical signal state 
is a simple but practically important problem. 
A typical setup of the optical phase estimation consists of a probe state, 
unitary (unknown) phase shifting, and the detection (and estimation) step
(Fig.~1). 
The precision of estimation is usually evaluated by a mean-square error 
$\mathrm{Var}[\hat{\phi}] = \mathrm{E}[(\hat{\phi}-\mathrm{E}[\hat{\phi}])^2]$ 
of the estimator $\hat{\phi}$. 
For unbiased estimator, it is well known that its ultimate lower bound 
is given by the quantum Cramer-Rao bound: 
\beq 
\mathrm{Var}[\hat{\phi}] \geq \frac{1}{MH(\phi)},
\label{eq:QCRB} 
\eeq
where $M$ is the number of data and $H(\phi)$ is 
the quantum Fisher information (QFI) \cite{BraunsteinCaves} 
which is a function of state $\rho(\phi)$. 

\begin{figure}[b]
\begin{center}
\includegraphics[width=0.9\linewidth]
{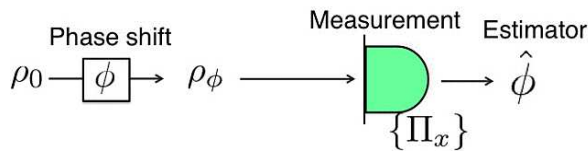}
\caption{Schematic of general phase estimation.
\label{setup}
}
\end{center}
\end{figure}

Two questions are particularly interesting 
and have attracted attentions so far: 1) what probe state maximizes 
$H(\phi)$ and 2) how to saturate the bound in realistic experimental setting. 
For the first question, it has been shown that highly nonclassical states, 
such as squeezed states　\cite{Monras,Furusawa,Berini} or NOON states　\cite{Giovannetti,NOON1,NOON2,Datta}, could beat the QFI of coherent state, 
which is often called the ``standard quantum limit'' scaling as $1/N$ where 
$N$ is the average photon number of the state. 
Moreover, ideally it reaches the so-called Heisenberg limit scaling $1/N^2$. 
However, it has been also revealed that these nonclassical states 
are very fragile to losses (unless using extremely nontrivial states) 
\cite{DDSLWBW09} that is unavoidable imperfection in real experiment. 
In this sense, for some practical applications in which one has to admit 
high losses, coherent state is still a useful option for phase sensing 
since coherent state preserves its coherence and purity even under high losses. In the following, we concentrate on phase estimation by coherent state probes.

The second question is related to the choice of the detection strategy. 
For a given system (i.e. given probe state and measurement), 
the lower bound of Var[$\hat{\phi}$] is determined by the classical 
Fisher information (FI) $F(\phi)$ and the QFI for a given probe state 
is defined as the maximum $F(\phi)$ over all possible quantum measurement 
\cite{BraunsteinCaves}. Thus, by definition, the following inequality holds for any 
given states:
\beq 
H(\phi) \geq F(\phi) ,
\label{eq:HF} 
\eeq
The question is to find the optimal measurement saturates this inequality preferably for any $\phi$.  
For coherent state, it is known that homodyne measurement can 
saturates (\ref{eq:HF}) around certain $\phi$ \cite{OlivaresParis}. 
This is a nice consequence but not always practical since the FI decrease 
quickly if $\phi$ is far from the specific point. 
Though globally optimal measurement is possible by adaptive homodyne detection 
with real time feedback system \cite{Hideo}, it may not be implementable 
for some applications, especially when the number of data is highly limited. 
In addition a finite bandwidth of the adaptive feedback operation causes 
some restrictions on the system parameters such as the repetition rate of 
source and detector. 
Therefore it is still worth to investigate the static measurement 
which could surpass the FI of the homodyne measurement 
in a wide range of $\phi$.

In this paper, we propose and experimentally demonstrate a novel and simple 
detection strategy for the coherent state phase estimation. 
Our detection scheme is inspired by the recent progress of 
``quantum receiver'' technology developed in the field of 
quantum optical communication. 
In optical communication at very low power regime, 
it is known that homodyne or heterodyne measurements are not optimal 
to minimize the error of discriminating modulated coherent state signals
\cite{Helstrom_book76_QDET}. 
In the last decade, practical quantum receiver configuration superior to 
homodyne and heterodyne receivers has been proposed for various set of signals 
\cite{Kennedy73,Dolinar73,Bondurant93,TakeokaSasakiLutkenhaus2006_PRL_BinaryProjMmt,Takeoka2008,Guha2011,izumi2012,izumi2013} 
and successfully demonstrated with the current technology 
\cite{CookMartinGeremia2007_Nature,Wittmann2008_PRL_BPSK,Tsujino2010_OX_OnOff,Tsujino2011_Q_Receiver_BPSK,Chen2012,Mueller2012_NJP,Becerra13,Becerra15}. 
Our basic idea is to apply these receivers --originally designed for state 
discrimination-- for a different purpose, i.e. the phase estimation. 
Specifically, we employ the simplest static receiver technique which we call 
the displaced photon counting 
\cite{Kennedy73,Takeoka2008,Tsujino2011_Q_Receiver_BPSK,Zhang12}, 
consisting of displacement operation and a photon detector 
(with no adaptive feedback). 
We show that in a wide range of $\phi$, our scheme works better than 
homodyne and heterodyne receivers in terms of FI. 
The concept is also demonstrated experimentally. 

The paper is organized as follows.
In Sec.~\ref{Sect:2}, we propose a novel detection scheme for 
the coherent state phase estimation and then analyze its performance 
in terms of the Fisher information formalism. 
Sec.~\ref{Sect:3} is devoted for the proof-of-principle experiment 
of the proposed scheme. 
Section \ref{Sect:4} concludes the paper. 
\section{Displaced-photon counting for phase estimation}\label{Sect:2}
In this section, after a brief reminder of the Fisher information formalism, 
we propose a simple detection scheme for phase estimation 
which we call the displaced-photon counting. 
We theoretically investigate its performance in terms of the FI and 
then compare it with that for other existing technologies such as homodyne 
and heterodyne detections. 

As mentioned in the introduction, the QFI sets the minimum bound of the variance for a given state.
For pure states, the QFI is simply given by \cite{OlivaresParis},
\beq 
H =4 \Delta K^2 ,
\label{eq:9} 
\eeq
where $ \Delta K^2 =\langle K^2 \rangle-\langle K\rangle^2$ is the fluctuation of the photon number of the state and $K=a^{\dagger} a$ is the photon number operator. 
$ \Delta K^2$ is invariant with respect to the phase shift $\phi$, and thus 
the QFI is simply determined by the initial state $\rho_0$.
Equation (\ref{eq:9}) allows us to explicitly calculate the QFI for coherent state, which simply turns out to be $H_{\mathrm{coh}}=4\alpha^2$.

The performance of a given system including both the state and the measurement can be theoretically analyzed by the FI \cite{Fisher}.
The FI is defined as 
\beq 
F(\phi)=\int \frac{1}{p(x|\phi)}(\frac{\partial p(x|\phi) }{\partial \phi})^2,
\label{eq:2} 
\eeq
where $p(x|\phi)$ is a conditional probability of obtaining an outcome $x$ for given phase shift $\phi$. This conditional probability is a function of the state$\rho_{\phi}$ and the measurement, described by a positive operator valued measure (POVM) $\{ \Pi_x  \}$ satisfying $\int dx \Pi_x=I $, as 
\beq 
p(x|\phi)=\tr{[\Pi_x \rho_{\phi}] }.
\label{eq:3} 
\eeq
Note that for discrete measurement observables such as photon numbers, the integration in (\ref{eq:2}) should be replaced with a sum of all measurement outcomes.
\begin{figure}[b]
\begin{center}
\includegraphics[width=0.9\linewidth]
{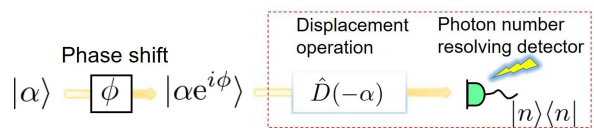}
\caption{
Schematic of the phase estimation for the coherent state measured by the displaced-photon counting.
\label{setupdis}
}
\end{center}
\end{figure}

Figure~\ref{setupdis} shows a simple schematic of the phase estimation of the coherent state with the displaced-photon counting.
Displacement operation $D(\beta)$ is a linear operation shifting the amplitude 
of coherent state $|\gamma\rangle$ as 
$D(\beta)\ket{\gamma}=\ket{\beta+\gamma}$.
In experiment it is realized by combining the signal light with a relatively strong local oscillator (LO) light on a beamsplitter with high transmittance. 
In our scheme, we implement $D(-\alpha)$ which converts the initial probe coherent state $\ket{\alpha}$ to a vacuum. 
This displacement is followed by a photon-number-resolving detector (PNRD). 
The measurement operator of the PNRD for $n$-photon outcome is given by \cite{BarnettPhillipsPegg1998},
\begin{equation}
\Pi_n =\mathrm{e}^{-\nu}\sum_{l=0}^{n}\sum_{k=n-l}^{\infty}\frac{\nu^{l}}{l!} C^{k}_{n-l} \eta^{n-l} (1-\eta)^{k-(n-l)}\proj{k},
\label{eq:8}
\end{equation}
where $C^{k}_{n-l}$ is the binomial coefficient, $\nu$ and $\eta$ are dark counts and detection efficiency, respectively.
For the initial state $|\alpha\rangle$ and a given phase shift $\phi$, 
the conditional probability of detecting $n$-photon is,
\begin{eqnarray}
p(n|\phi)&=& \xi \frac{ \bigl( 2\eta \alpha^2(1-\cos{\phi})+\nu \bigr) ^n}{n!} \mathrm{e}^{-2\eta\alpha^2(1- \cos{\phi})-\nu} 
\nonumber
\\
&& +
2(1-\xi) \frac{(\eta \alpha^2+\nu)^n}{n!} \mathrm{e}^{-\eta \alpha^2 -\nu}.
\label{eq:13} 
\end{eqnarray}
In the ideal case, where $\alpha=\beta$, $\nu=0$, $\eta=1$ and 
the perfect interference visibility ($\xi=1$) \cite{Takeokavisi},
an analytical form of the FI is obtained as 
\beq
F_{\mathrm{dis}}(\phi) =2 \alpha^2 (1+\cos^2{\phi}).
\label{eq:14} 
\eeq
This should be compared with the FIs of the conventional homodyne and heterodyne detections \cite{OlivaresParis,gaussian_review}. 
The POVM of the homodyne detection is given by a set of projectors onto quadrature bases $\{ \Pi_p=\ket{p}\bra{p} \}$ and it implies,
\beq 
F_{\mathrm{hom}}(\phi)=4 \alpha^2 \cos^2{\phi}.
\label{eq:11} 
\eeq
The heterodyne detection is composed of a balanced beam splitter followed by two homodyne detectors for conjugate quadrature components. 
Its POVM is given by a set of coherent states 
$\frac{1}{\pi}\{ \Pi_\beta = |\beta\rangle\langle\beta| \}$. 
Though the heterodyne measurement can measure two quadrature amplitudes 
simultaneously, its performance is limited by the unwanted vacuum fluctuation 
input from the unused port of the balanced beam splitter. 
Thus the QFI is limited to be 
\beq 
F_{\mathrm{het}}(\phi)=2 \alpha^2.
\label{eq:12} 
\eeq

\begin{figure}[h]
\begin{center}
\includegraphics[width=0.9\linewidth]
{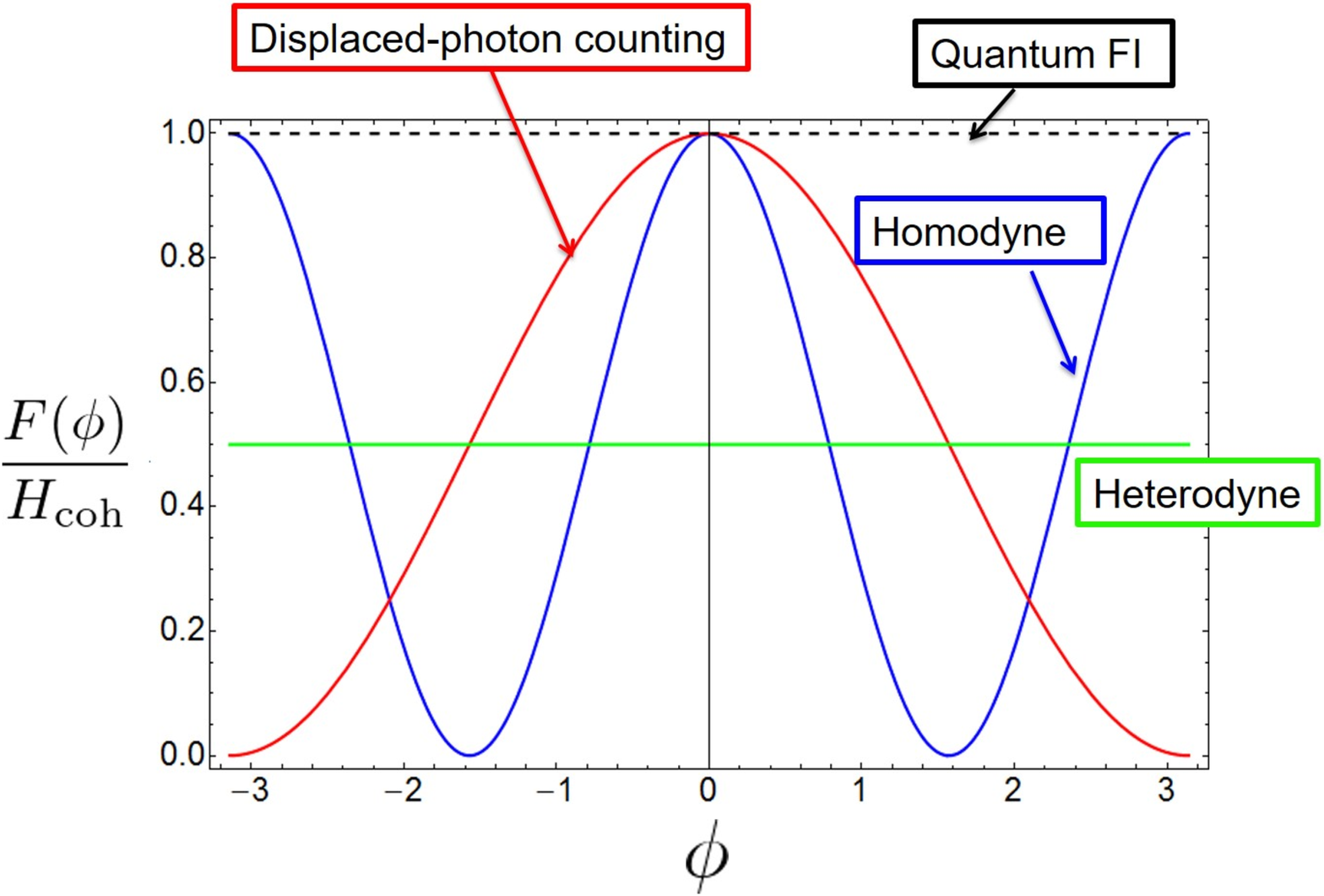}
\caption{Fisher Information for the displaced-photon counting, the homodyne detection and the heterodyne detection.
Vertical axis is normalized by the Quantum Fisher Information
 of the coherent state.
 Thus maximum value is $1$ where the FI corresponds to the QFI.  
\label{Fisher Information}
}
\end{center}
\end{figure}

In Fig.~\ref{Fisher Information} we plot the FIs for three-type measurements, the displaced-photon counting, the homodyne and the heterodyne detection.
The homodyne detection's FI can reach the QFI at a local point $\phi=0, \pm\pi$.
However, it decreases rapidly as the phase $\phi$ shifts 
from this local points whereas the heterodyne detection's FI is constantly a half of the QFI. 
We observe that the displaced-photon counting also reaches the QFI at a local phase point $\phi=0$ and filling the gap between the homodyne and the heterodyne 
in the sense that it is better than both in a wide region around $\phi=0$. 
This might be useful in some applications where the sample's phase is known to be around $\phi=0$ but has relatively wide fluctuation around there. 
\begin{figure}[h]
\centering
\subfigure
{
\includegraphics[width=0.95\linewidth]
{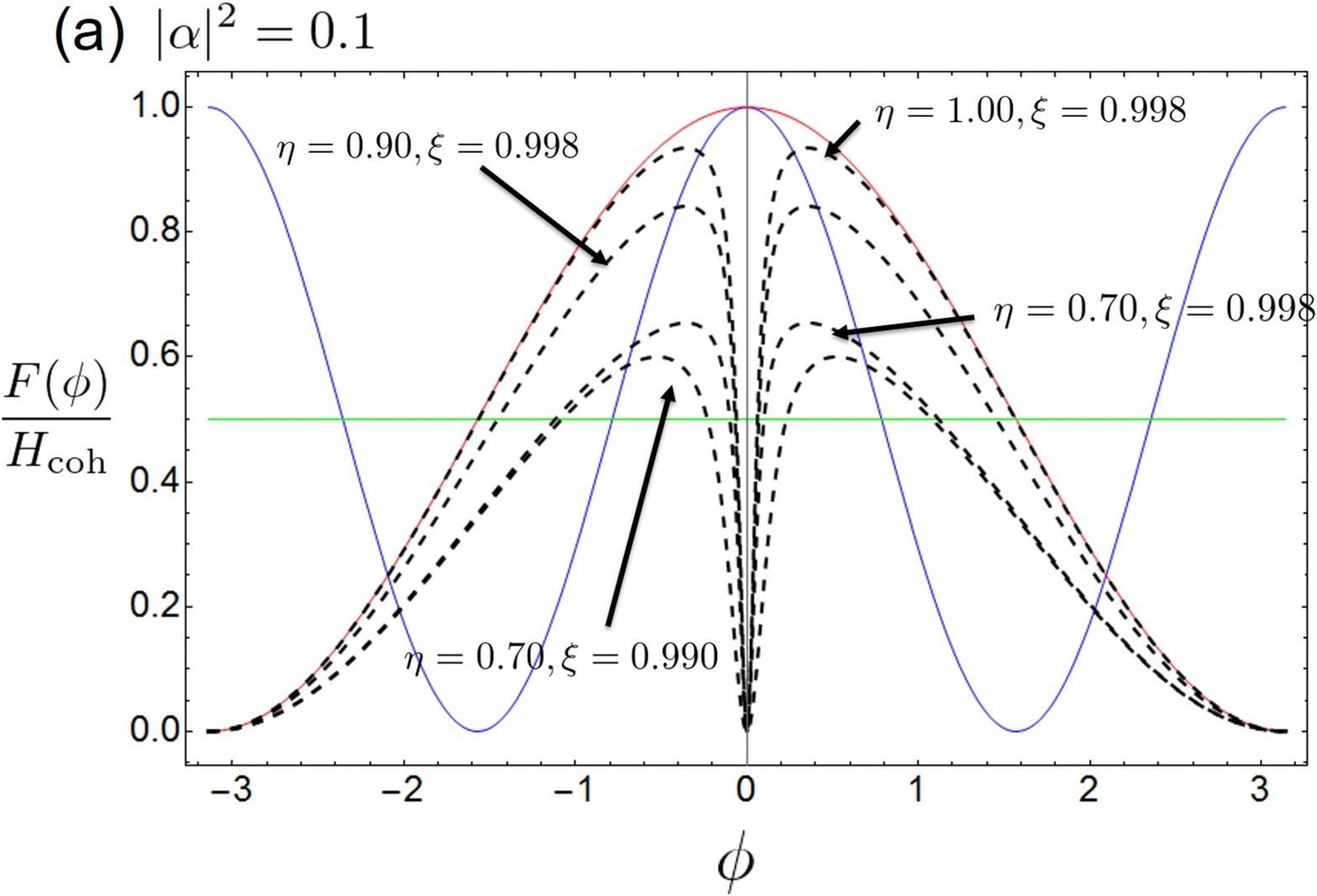}
\label{mean0.1}
}
\subfigure
{
\includegraphics[width=0.95\linewidth]
{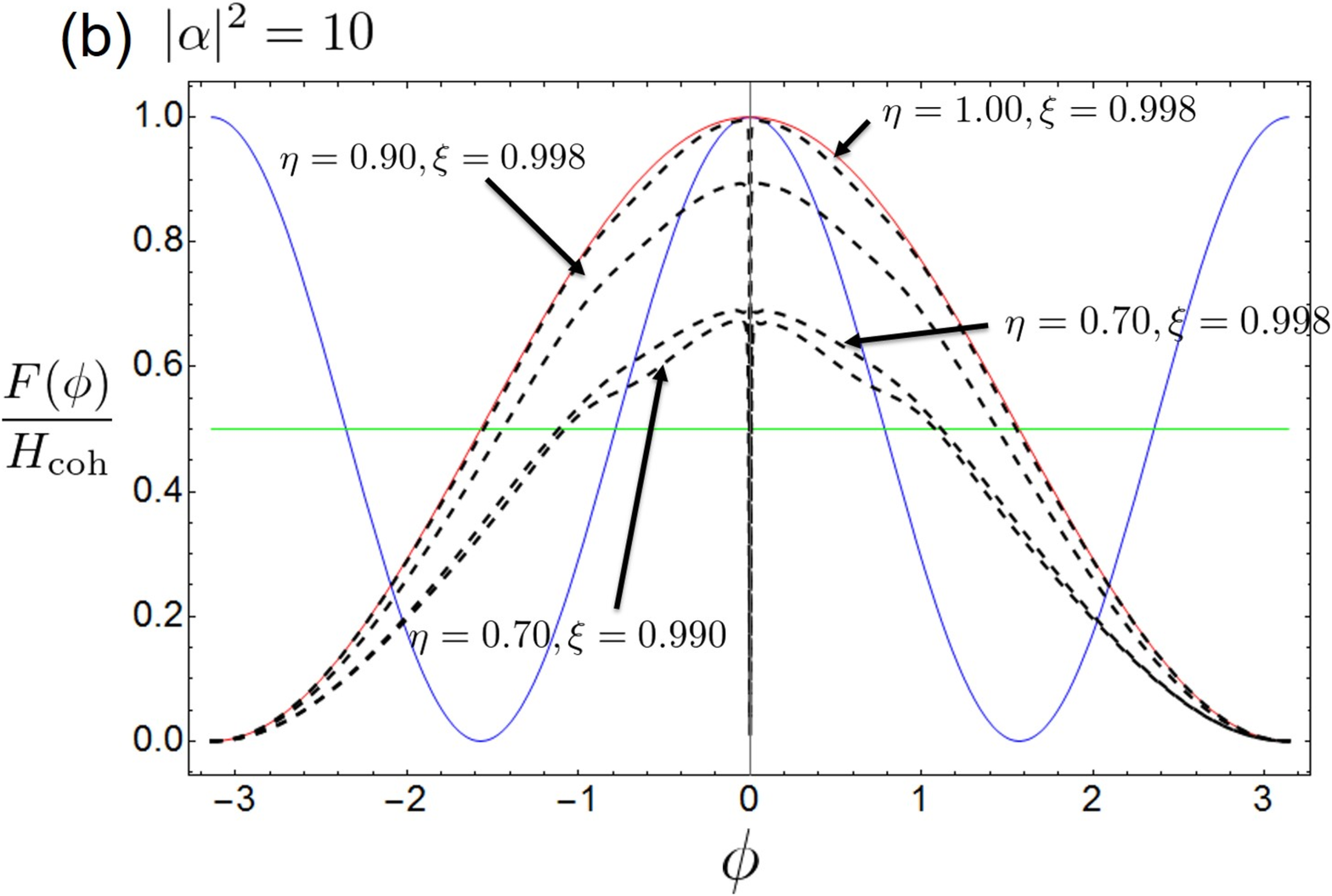}
\label{mean10}
}
\caption{
Normalized FI of the displaced-photon counting with the various imperfect conditions and fixed dark count $10^{-5}$ counts per pulse.
The mean photon number of the signal state is
(a)\, $\abs{\alpha}^2=0.10$
and
(b)\, $\abs{\alpha}^2=10$.
}
\label{FIwithloss}
\end{figure}
\begin{figure*}[t]
\centering
{
\includegraphics[width=0.6\linewidth ]{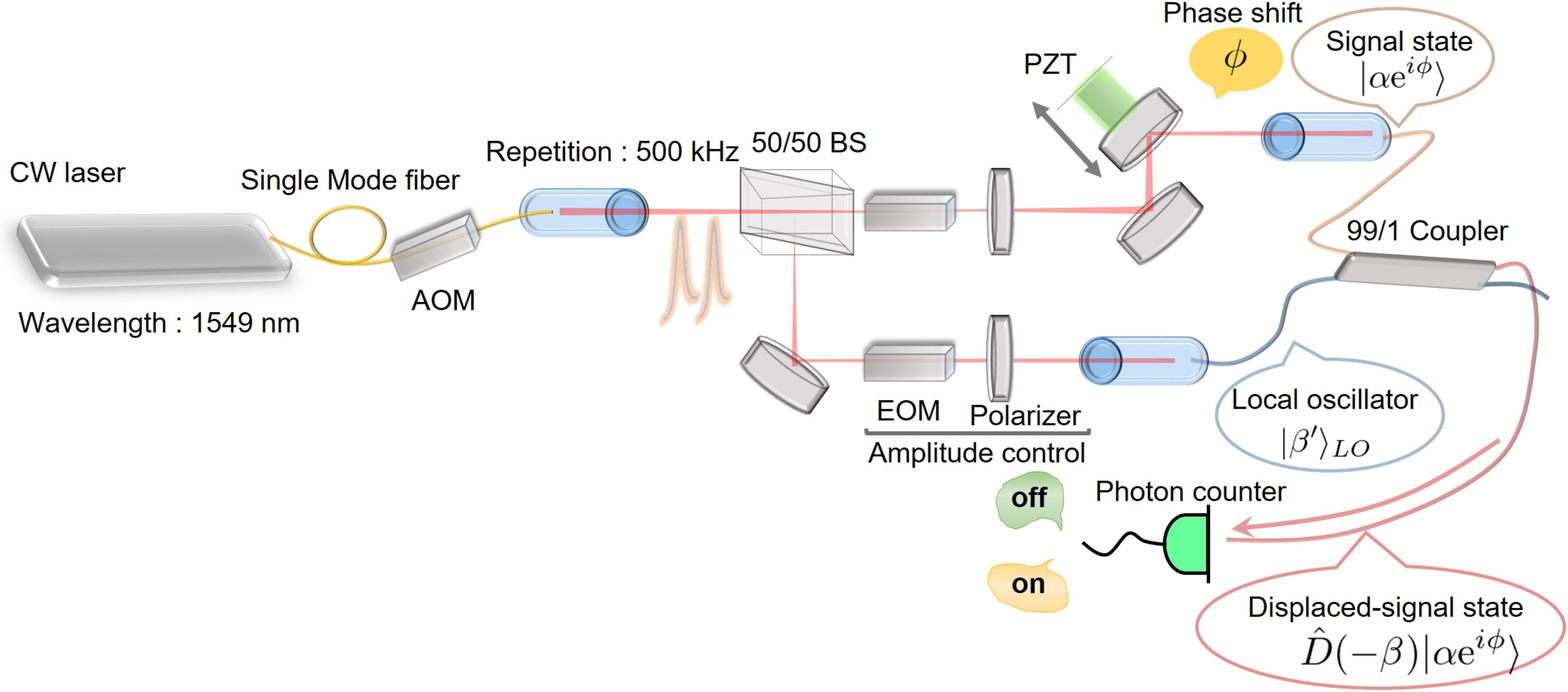}
}
\caption{
Experimental setup.
AOM : Acousto-optic modulator, 
BS :  Beam splitter, 
EOM : Electro-optic modulator, 
PZT : Piezo transducer.}
\label{expsetup}
\end{figure*}
The FI of the displace-photon counting with the imperfections are illustrated in Fig.~\ref{FIwithloss}.
The performance of the displaced-photon counting is degraded because of the non unit detection efficiency.
Furthermore, in case of the small mean photon number Fig.~\ref{FIwithloss} (a), the phase insensitive noises such as the imperfect visibility and the dark counts cause high degradation of the FI around $\phi=0$.
This is because the phase insensitive terms in the Eq.~(\ref{eq:13}) become dominant and make the conditional probability less sensitive to the phase shift.
In Fig.~\ref{FIwithloss} (b), we show the FI of the displaced-photon counting with large coherent amplitude $\abs{\alpha}^2=10$.
The dark count noise and the imperfect visibility do not make a critical contribution to the conditional probability for the large signal amplitude conditions (see around $\phi=0$). Thus the scheme is robust against the phase insensitive noise for larger $\abs{\alpha}^2$ than the weaker probe in Fig.~\ref{FIwithloss} (a).

In real experiment, the measurement outcome from the detector should be post-processed to estimate the phase value. As a post-processing algorithm, we choose a standard Bayesian strategy, which is known to saturate the (classical) Cramer-Rao bound  $\mathrm{Var}[\hat{\phi}] \geq 1/MF(\phi)$ for large enough $M$ \cite{OlivaresParis}. 

Suppose $\{ n_k \} = \{ n_1,  n_2, \cdots , n_k \}$ is a set of photon numbers observed by the detector after $k$ measurements. 
Then a ${\it posteriori}$ probability $P(\phi | \{ n_k \})$ of $\phi$ given 
$\{ n_k \}$ is obtained from a relation,
\beq 
p(\{ n_k \})P(\phi | \{ n_k \}) = p(\phi)P( \{ n_k \} |\phi).
\label{eq:14} 
\eeq
The prior probability distribution $p(\phi)$ is assumed to be uniform distribution in our estimation and $p(\{ n_k \})=\int_{0}^{\pi} p(\phi)P( \{ n_k \} |\phi)d\phi$ is the prior probability of observing $\{ n_k \}$. 
The latter can be calculated from Eq.~(\ref{eq:13}) and the relation: 
\beq 
P(\{ n_k \} |\phi) = \prod_{i=1}^{k} p(  n_i |\phi).
\label{eq:15} 
\eeq
Combining Eqs.~(\ref{eq:14}) and (\ref{eq:15}) and the uniform $p(\phi)$, 
we can explicitly derive the form of $P(\phi | \{ n_k \})$. 
Then an expectation value of the estimator and its variance 
for experimentally measured $\{ n_k \}$ 
are evaluated as,
\beqa
\hat{\phi}&=&\int \phi P(\phi | \{ n_k \}) d\phi, 
\label{eq:17} 
\\
\mathrm{Var}[\hat{\phi}] &=& \int  (\hat{\phi}-\phi )^2  P(\phi | \{ n_k \}) d\phi.
\label{eq:18} 
\eeqa
\section{Experiment}\label{Sect:3}
Figure \ref{expsetup} shows our experimental setup.
Continuous wave laser at 1549 nm is modulated to a sequence of optical pulses with repetition rate 900 kHz and pulse width 100 nm by an acousto-optic modulator (AOM).
The optical pulse is sent into a Mach-Zehnder interferometer in which
amplitudes of the optical lights are independently controlled by a set of an electro-optic modulator (EOM) and a polarizer.
In an optical path for the signal coherent state, a piezo transducer produces the optical phase shift 
$\phi$ to be estimated.
The signal state with the coherent amplitude $\alpha$ is defined after a fiber coupling
and displaced by combining the LO light on an asymmetric fiber coupler with transmittance $\tau=0.99$.
We achieve the visibility $\xi =0.993 $ for the displacement operation.
In the experiment, instead of a PNRD, we use a superconducting nanowire single photon detector (SNSPD), which only discriminate if the photons exist or not \cite{SSPD, SSPD2}. 
The performance gap between the SNSPD and the ideal PNRD are huge when 
$\alpha$ is large. However, in this proof-of-principle experiment, we choose $\alpha \ll 1$ such that the performance gap between them is in principle negligible because of the extremely small probability of having more than one photon per pulse. 
We calibrate the optical power of the laser light by using a well calibrated power meter and then insert a well-calibrated attenuator which reduces the power of the optical light up to single photon level.
The power meter is replaced with the SSPD and, by comparing the expected mean photon number after the attenuation and the detected mean photon number, we obtain the total detection efficiency of our system $\eta =60.2 \pm 0.4 \% $ \cite{TES2}.
Electrical signals from the SNSPD are first transmitted to SR400 gated photon counter (Stanford Research Systems)
which enables to reduce the dark count to $\nu = 1.13  \times 10^{-4} $ counts/pulse by gating the detection window synchronized with the AOM.
The actual phase shift is inferred from $900\times 10^{3}$ points counted by the SR400 without time information
and we simultaneously monitor the electrical signals using oscilloscope up to $1\times 10^{4}$ points with time information which are shown as the experimental results in this paper.

\begin{figure}[t]
\centering
\subfigure
{
\includegraphics[width=1\linewidth]
{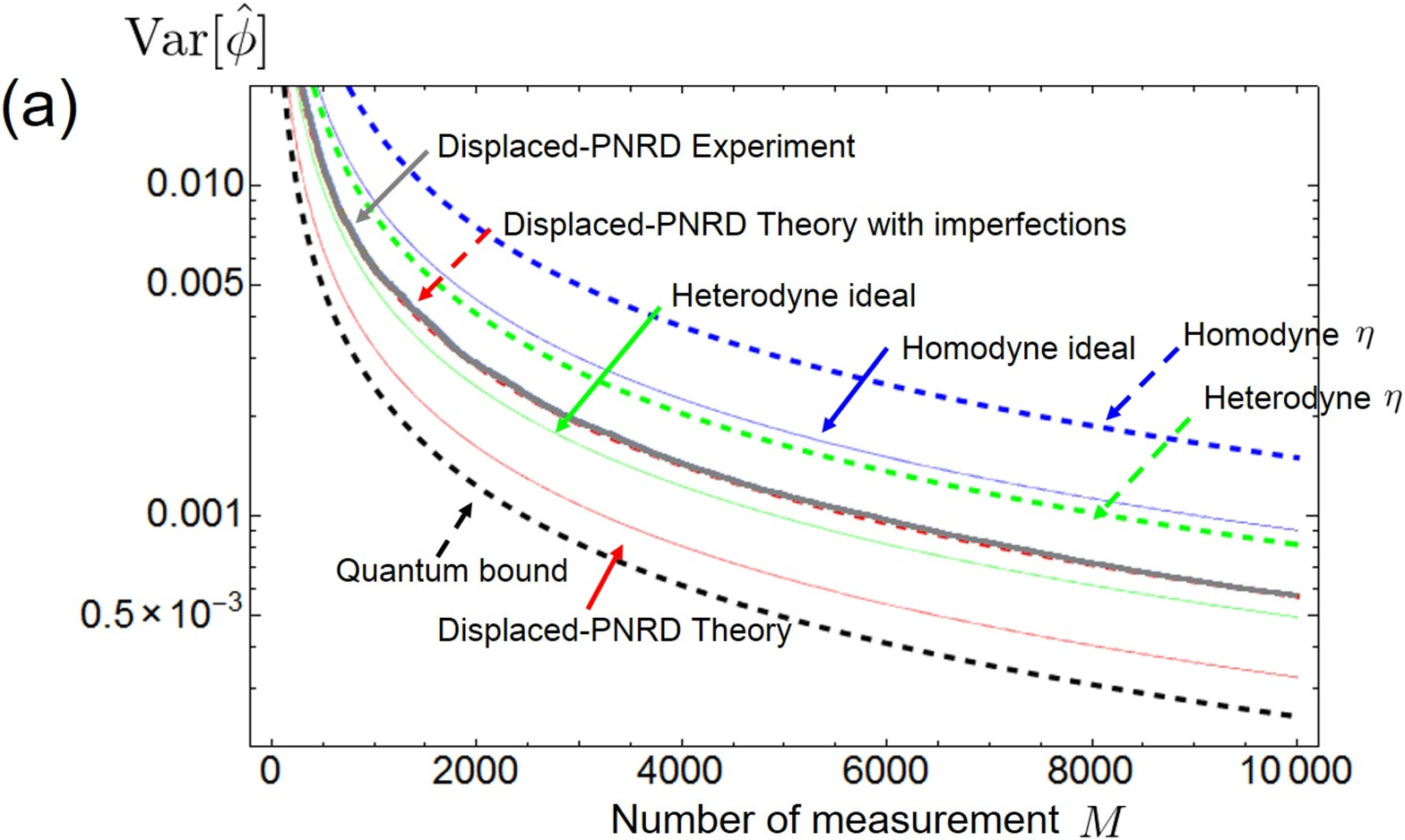}
\label{var1}
}
\subfigure
{
\includegraphics[width=0.85\linewidth]
{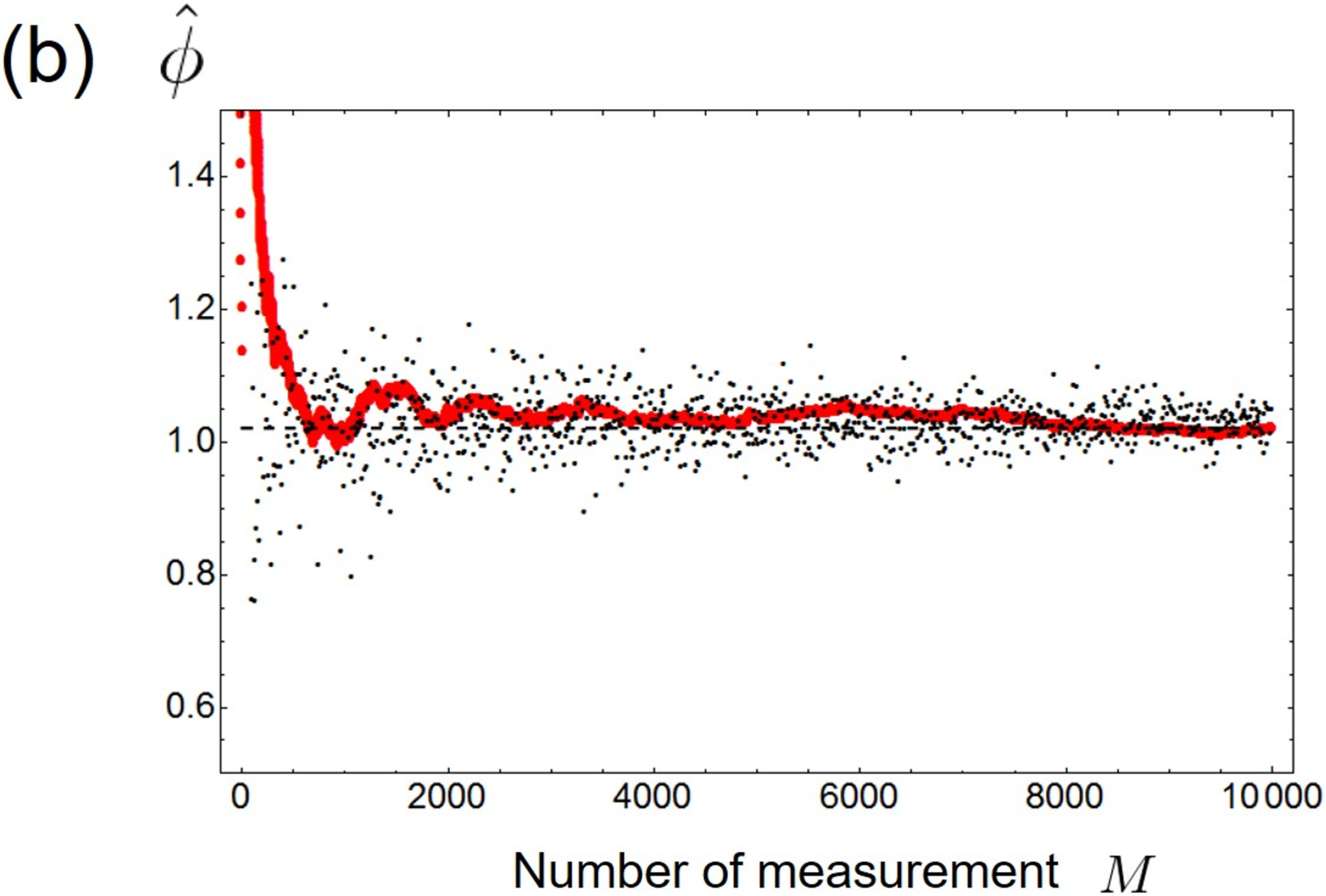}
\label{mean1}
}
\caption{Experimental results for 
(a)\, the variance of the estimator,
(b)\, the expectation value of the estimator (red dot) with 
the data points generated from numerical simulation (black dot).
The phase shift value is fixed to $\phi \sim 1.00$ and the parameters for the experiment are $\abs{\alpha}^2= 0.100,  \abs{\beta}^2 =  0.101, \xi =0.993, \eta =60.2 \%$ and $\nu = 1.13 \times 10^{-4}$.
}
\label{phi1}
\end{figure}

We fix the intensities of the signal state $\abs{\alpha}^2= 0.100$ and the displacement operation $\abs{\beta}^2 =  0.101$ with the uncertainty $1\%$.
Figure \ref{phi1}
depicts the experimental results for (a) the variance of the estimator and (b) the expectation value when the optical phase shift is set to $\phi \sim1.00$.
In Fig.~\ref{phi1}(a), we compare the experimentally evaluated variance of the estimator (gray) with the homodyne detection (blue dashed), the heterodyne detection (green dashed), the displaced-photon counting with (red dashed) and without (red solid) experimental imperfections and quantum bound derived from the QFI (black dashed).
We observe a good agreement between the experimental results and the theoretical predictions.
The expectation value estimated from $M$-measurement outcomes is shown by red dots in Fig.~\ref{phi1}(b) as a function of the number of measurements $M$.
Black dots in the Fig.~\ref{phi1}(b) are independently generated by numerical simulation therefore reflect statistical fluctuation.
The expectation value well locates at the actual value of the phase shift (black dashed line) except for the small $M$.
Even though our photon counter and non-unit visibility degrade the performance of the displaced-photon counting, our detection strategy still shows the performance overcoming the homo- and hetero-dyne detections for the specific phase conditions.

\begin{figure}[t]
\centering
\subfigure
{
\includegraphics[width=0.9\linewidth]
{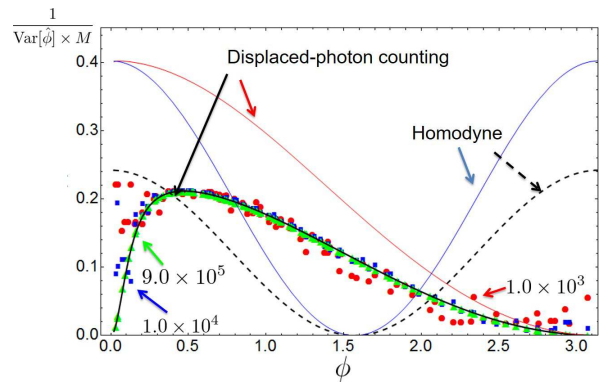}
\label{InverseFI}
}
\caption{
Inverse of the experimentally obtained variance of the estimator multiplied the number of measurements $M$.
$M$ is set to $1.0\times 10^{3}$ (red circle), $1.0\times 10^{4}$ (blue square) and $9.0\times 10^{5}$ (green triangle).
Black solid and dashed lines represent the theoretical FI of the displaced-photon counting and the homodyne detection for the experimental condition $\eta=0.602, \nu = 1.13  \times 10^{-4}$ and $\xi =0.993$.
Red and black solid lines are the theoretical FI of the displaced-photon counting and the homodyne detection in ideal case.
}
\label{FI_Bias}
\end{figure}
The Bayes' theorem guarantees the saturation of the Cramer-Rao inequality for the unbiased estimator if the sample size is infinity.
To verify the convergence property for the finite number of samples,
we plot the inverse of the variance multiplied the number of measurements $M$ in Fig.~\ref{FI_Bias} as a function of the optical phase shift $\phi$.
Circle (red), square (blue) and triangle (green) plot correspond to the number of measurements $M=1.0\times 10^{3}, 1.0\times 10^{4}$ and $9.0\times 10^{5}$ respectively.
The ideal FIs without any imperfections for the displaced-photon counting and the homodyne detection are shown by red and blue solid lines, respectively.
The experimental data well coincide with the FI of the displaced-photon counting for the experimental condition (solid black line).
This coincidence indicates that the Cramer-Rao inequality is saturated with the given number of measurements.
However, 
the discrepancy between the FI and the experimental results becomes apparent when $\phi$ is extremely small or close to $\pi$.
The performance overcoming the theoretical FI could occur in the finite samples case because of the statistical randomness of the data acquisition.
Note that the variance evaluated from sufficiently large number of samples or the mean of the variance obtained from independent trials in the same condition satisfies the Cramer-Rao inequality.

\section{Conclusion}\label{Sect:4}
We proposed and experimentally demonstrated a simple optical phase estimation detector which we call the displaced-photon counting. 
The detector configuration was inspired by suboptimal receiver 
for a signal discrimination in quantum optical communication and 
its application to phase estimation has been examined 
in both theory and experiment. 
The theoretical results showed that the displaced-photon counting exhibits near optimal performance 
in a wider range of the phase value $\phi$ than the conventional homodyne detection.
The features of the displaced-photon counting offer advantage especially when 
the optical phase shift is unknown but supposed to be located around $\phi=0$.
The proof-of-principle of our detection scheme is demonstrated by using the SNSPD as a photon counting device. Though our SNSPD is not ideal (non-unit detection efficiency ($60.2\%$), finite dark counts ($1.13 \times 10^{-4}$ per pulse), 
and indistinguishability of photon numbers), our result still overcome the ideal limit of homodyne and heterodyne detections around $\phi=0$. 
Also the results well agree with our theoretical predictions. 
That is we experimentally observed the saturation of the Cramer-Rao bound 
via the estimation based on the Bayesian strategy.

There are several interesting future directions. 
The first possible direction is installation of the PNRD as the photon counter.
Though our proof-of-principle experiment with the on/off detector overcome the homo- and hetero-dyne limits in small mean photon number, it does not work for stronger probes e.g. few photons per pulse. Moreover, as theoretically show in Sec.\ref{Sect:2}, the PNRD is robust against the phase insensitive noise by dark counts and the imperfect visibility.
The efficiency of the detector is also critical issue that degrades the performance of our scheme and required to be almost unity to observe the expected performance.
One of the candidates of the desired PNRD satisfying above conditions is transition edge sensors \cite{TES, TES2}.
Another direction is its applications. 
Our phase estimation strategy using coherent state provides benefit for an optical metrology especially when signal light is weak, losses are not negligible and sample size is not enough large to utilize an adaptive measurement.
Finally an interesting future problem is the application of the phase estimation strategy using the displaced-photon counting into the coherent communication scenario.
Although the task in communication is to discriminate the discretely encoded signals with minimum error, the phase estimation is also very important since in practice the receiver has to track the sender's reference frame. 
The phase estimation of the binary encoded coherent states in the binary optical communication is discussed in \cite{Bina}. 
Our detailed analysis could be useful for designing a future optical communication system with extremely weak signal and practical phase tracking.

\begin{acknowledgements}
We acknowledge J. P. Dowling and K. P. Seshadreesan for fruitful discussion.
This work was supported from
Open Partnership Bilateral Joint Research Projects
(JSPS) and ImPACT Program of Council for Science,
Technology and Innovation (Cabinet Office, Government of Japan).
\end{acknowledgements}



\end{document}